\documentclass[twocolumn,aps,prc,superscriptaddress,showpacs,floatfix]{revtex4}
\usepackage{graphicx}
\usepackage{dcolumn}
\usepackage{bm}

\usepackage[normalem]{ulem} 
\usepackage[dvips]{color} 
\renewcommand\sout{\bgroup \color{red} \ULdepth=-.5ex \ULset}

\begin{document}

\preprint{}
\title{Neutron-proton bremsstrahlung from intermediate energy heavy-ion
reactions as a probe of the nuclear symmetry energy?}
\author{Gao-Chan Yong}
\affiliation{Department of Physics, Texas A\&M
University-Commerce, Commerce, Texas 75429, USA}
\affiliation{Institute of Modern Physics, Chinese Academy of
Science, Lanzhou 730000, China}\affiliation{Graduate School,
Chinese Academy of Science, Beijing 100039, P.R. China}
\author{Bao-An Li}
\affiliation{Department of Physics, Texas A\&M
University-Commerce, Commerce, Texas 75429, USA}
\author{Lie-Wen Chen}
\affiliation{Department of Physics, Texas A\&M
University-Commerce, Commerce, Texas 75429, USA}
\affiliation{Institute of Theoretical Physics, Shanghai Jiao Tong
University, Shanghai 200240, China}\affiliation{Center of
Theoretical Nuclear Physics, National Laboratory of Heavy Ion
Accelerator, Lanzhou 730000, China}

\begin{abstract}
Hard photons from neutron-proton bremsstrahlung in intermediate
energy heavy-ion reactions are examined as a potential probe of
the nuclear symmetry energy within a transport model. Effects of
the symmetry energy on the yields and spectra of hard photons are
found to be generally smaller than those due to the currently
existing uncertainties of both the in-medium nucleon-nucleon cross
sections and the photon production probability in the elementary
process $pn\rightarrow pn\gamma$. Very interestingly,
nevertheless, the ratio of hard photon spectra $R_{1/2}(\gamma)$
from two reactions using isotopes of the same element is not only
approximately independent of these uncertainties but also quite
sensitive to the symmetry energy. For the head-on reactions of
$^{132}Sn+^{124}Sn$ and $^{112}Sn+^{112}Sn$ at $E_{beam}/A=50$
MeV, for example, the $R_{1/2}(\gamma)$ displays a rise up to 15\%
when the symmetry energy is reduced by about 20\% at
$\rho=1.3\rho_0$ which is the maximum density reached in these
reactions.
\end{abstract}

 \maketitle
\date{\today}

The density dependence of the nuclear symmetry energy
$E_{sym}(\rho)$ is important for understanding many interesting
questions in both nuclear physics and
astrophysics\cite{lat01c,steiner05a}. In particular, the reaction
dynamics and many observables of heavy-ion reactions are strongly
influenced directly by the corresponding nuclear symmetry
potential\cite{ireview,ibook,baran05}. In turn, these observables
can be used as effective probes of the symmetry energy
$E_{sym}(\rho)$. Indeed, considerable progress has been made
recently in constraining the $E_{sym}(\rho)$ at sub-normal
densities using heavy-ion
reactions\cite{tsang04,chen05,li05,shetty07}. However, the
behavior of the nuclear symmetry energy at supranormal densities
remains rather illusive. A quick review of the current situation
of the field\cite{chen07} reveals that essentially all known
probes of the $E_{sym}(\rho)$ in heavy-ion reactions are hadronic
in nature. These probes, especially if used for studying the high
density behavior of the $E_{sym}(\rho)$, inevitably suffer from
distortions due to the strong interactions in the final state.
Although selections of some especially delicate observables in
certain kinematic/geometrical regions, such as, the neutron/proton
ratio of squeezed-out nucleons perpendicular to the reaction
plane\cite{yong07}, are promising in reducing effects of the final
state interaction, ideally one would like to have more clean ways
to study the symmetry energy especially at supranormal densities.
In this regard, it is interesting to note that the
parity-violating electron scattering has been proposed to measure
more precisely the size of the neutron-skin in
$^{208}$Pb\cite{hor01}. The latter measured previously in
experiments using hadronic probes was shown recently in several
calculations to be proportional to the slope of the symmetry
energy at and/or slightly below the normal
density\cite{steiner05a,brown00,typel01,furn02,kara02,stone03,steiner05b,chen05nskin}.
Since only electromagnetic interactions are involved in electron
scatterings the proposed experiment PREX is theoretically the most
clean probe known so far for probing the low density behavior of
the symmetry energy\cite{prex1,prex2}. However, practically it is
very challenging since the probability of obtaining
parity-violating electron scattering events in the experiment is
extremely small. The long waited data is hopefully coming soon.
Similar to electrons, photons interact with nucleons only
electromagnetically. Once produced they escape almost freely from
the nuclear environment in nuclear reactions. We notice that soft
photons from giant dipole resonances in heavy-ion reactions have
been shown within a semiclassical molecular dynamics model to be
quite sensitive to the symmetry potential term in the
nucleon-nucleon interaction\cite{giu06}. Can one use hard photons
from intermediate energy heavy-ion reactions to extract
information about the $E_{sym}(\rho)$ especially at supranormal
densities? This is a question that one of the present authors was
repeatedly asked recently by several experimentalists on several
occasions\cite{wie}. To answer this important question, we report
here results of the first exploratory study on using hard photons
from neutron-proton bremsstrahlung in intermediate energy
heavy-ion reactions as a probe of the $E_{sym}(\rho)$.

Hard photon production in heavy-ion reactions at beam energies
between about 10 and 200 MeV/A had been extensively studied both
experimentally and theoretically during the last two decades, see,
e.g., refs.\cite{bertsch88,nif90,cassrp} for a comprehensive
review. Indeed, very interesting physics has been obtained from
analyzing data taken by several experimental collaborations. For
instance, the TAPS collaboration carried out a series of
comprehensive measurements at various experimental facilities
(GSI, GANIL, KVI) studying in detail the properties (energy
spectra, angular distributions, total photon multiplicities
di-photon correlation functions, etc), of hard photons in a large
variety of nucleus-nucleus systems in the range of energies
spanning $E_{lab}\approx 20-200$ MeV/nucleon\cite{david}. They
used those bremsstrahlung photons as a tool to study the nuclear
caloric curve, the dynamics of nucleon-nucleon interactions, as
well as the time-evolution of the reaction process before nuclear
break-up\cite{TAPS}. Theoretically, it was concluded that the
neutron-proton bremsstrahlungs in the early stage of the reaction
are the main source of high energy $\gamma$ rays. Within the
cascade and Boltzmann-Uehling-Uhlenbeck (BUU) transport models it
was demonstrated clearly that the hard photons can be used to
probe the reaction dynamics leading to the formation of dense
matter\cite{bertsch86,ko85,cassing86,bau86,stev86}. However,
effects of the nuclear Equation of State (EOS) on the hard photon
production was found small\cite{ko87}. While these reaction models
were able to reproduce all qualitative features of the
experimental data, normally the quantitative agreement is within
about a factor of 2. One of the major uncertainties is the input
elementary $pn\rightarrow pn\gamma$ probability $p_{\gamma}$ which
is still rather model
dependent\cite{nif85,nak86,sch89,gan94,tim06}. It was noticed
earlier that the few existing data for the $pn\rightarrow
pn\gamma$ process can be described reasonably well by the
available models usually within a factor of 2\cite{cassrp}.
Looking forward enthusiastically, we mention here that the very
recent systematic measurements of the $pn\rightarrow pn\gamma$
cross sections with neutron beams up to 700 MeV at Los Alamos have
the potential to improve the situation significantly in the near
future\cite{saf07}.

Since the photon production probability is so small, i.e., only
one in roughly a thousand nucleon-nucleon collisions produces a
photon, a perturbative approach has been used in all dynamical
calculations of photon production in heavy-ion reactions at
intermediate energies\cite{bertsch88,cassrp}. In this approach,
one calculates the photon production as a probability at each
proton-neutron collision and then sum over all such collisions
over the entire history of the reaction. As discussed in detail
earlier in ref.\cite{cassrp}, the cross section for neutron-proton
bremsstrahlung in the long-wavelength limit separates into a
product of the elastic $np$ scattering cross section and a
$\gamma$-production probability. The probability is often taken
from the semiclassical hard sphere collision
model\cite{bertsch88,nif90,cassrp}. The double differential
probability, ignoring the Pauli exclusion in the final state, is
given by
\begin{eqnarray}\label{coss}
\frac{d^{2}N}{d\varepsilon_{\gamma}d\Omega_{\gamma}}&=&\frac{e^{2}}{12\pi^{2}\hbar
c}\times\frac{1}{\varepsilon_{\gamma}}(3\sin^{2}\theta_{\gamma}\beta_{i}^{2}+2\beta_{f}^{2})\nonumber \\
&=&6.16\times10^{-5}\times\frac{1}{\varepsilon_{\gamma}}(3\sin^{2}\theta_{\gamma}\beta_{i}^{2}+2\beta_{f}^{2}),
\end{eqnarray}
where $\theta_{\gamma}$ is the angle between the incident proton
direction and the emission direction of photon; and $\beta_i$ and
$\beta_f$ are the initial and final velocities of the proton in
the proton-neutron center of mass frame. The above equation was
obtained from modifying the original semi-classical Jackson
formula\cite{jackson} to allow for energy conservation in the
$\gamma$-production process\cite{cassing86,bau86}. Integrating
Eq.\ (\ref{coss}) over the photon emission angle, one obtains the
single differential probability
\begin{equation}\label{intc}
p^a_{\gamma}\equiv\frac{dN}{d\varepsilon_{\gamma}}
=1.55\times10^{-3}\times\frac{1}{\varepsilon_{\gamma}}(\beta_{i}^{2}+\beta_{f}^{2}).
\end{equation}
We notice that other expressions derived theoretically involving
more quantum-mechanical effects exist in the literature, see,
e.g., \cite{nif85,nak86,sch89,gan94,tim06}. Without passing any
judgement on these theories, to simply evaluate influences of the
elementary $pn\rightarrow pn\gamma$ probability on photon
production in heavy-ion reactions, for a comparison we thus also
use the prediction of the one boson exchange model by Gan et
al.\cite{gan94}
\begin{equation}\label{QFT}
p^b_{\gamma}\equiv\frac{dN}{d\varepsilon_{\gamma}}=2.1\times10^{-6}\frac{(1-y^{2})^{\alpha}}{y},
\end{equation}
where $y=\varepsilon_{\gamma}/E_{max}$, $
\alpha=0.7319-0.5898\beta_i$, and $E_{max}$ is the energy
available in the center of mass of the colliding proton-neutron
pairs.

\begin{figure}[th]
\begin{center}
\includegraphics[width=0.45\textwidth]{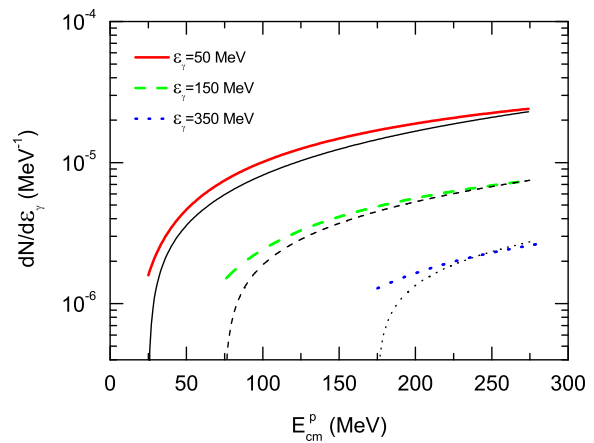}
\end{center}
\caption{(Color online) The single differential probability as a
function of proton kinetic energy in the proton-neutron center of
mass frame for the production of photons at energies of $50$,
$150$ and $350$ MeV, respectively. The lines with higher values
are results calculated with the semi-classical Eq. (\ref{intc})
while the ones with lower values are obtained by using the
quantum-mechanical Eq. (\ref{QFT}).} \label{cross}
\end{figure}

The single differential probability $p^a_{\gamma}$ and
$p^b_{\gamma}$ from the two models are shown in Fig.\ \ref{cross}
as a function of proton kinetic energy in the proton-neutron
center of mass frame for the production of photons at energies of
$50$, $150$ and $350$ MeV, respectively. It is seen that the two
models give quite similar but quantitatively different results
especially near the kinematic limit where the $p^a_{\gamma}$ is
significantly higher than the $p^b_{\gamma}$, as noticed already
in ref.\cite{gan94}. Moreover, the very small magnitude of the
photon production probability shown here justifies the use of the
perturbation method.

In most of the previous calculations for hard photon production
using transport models, a constant nucleon-nucleon cross section
of 30mb to 40mb was normally used\cite{ko85,bau86,cassing86,ko87}.
In this study, we use the IBUU04 transport model\cite{li04a} where
an isospin-dependent in-medium nucleon-nucleon (NN) cross section
\begin{equation}
\sigma^{medium}_{NN}=\sigma^{free}_{NN}(\frac{\mu^*_{NN}}{\mu_{NN}})^2
\end{equation}
was implemented\cite{li05}. In the above, the $\mu^*_{NN}$ and
$\mu_{NN}$ are the in-medium and free-space reduced NN mass.
Because of the neutron-proton effective mass splitting due to the
momentum dependence of the isovector potential, the scaling factor
$ (\frac{\mu^*_{NN}}{\mu_{NN}})^2$ induces a significant
modification to the relative cross sections of $np$, $nn$ and $pp$
as discussed in detail in ref.\cite{li05}. The energy and isospin
dependent free-space NN cross section $\sigma^{free}_{NN}$ are
taken from the experimental data.

\begin{figure}[th]
\begin{center}
\includegraphics[width=0.45\textwidth]{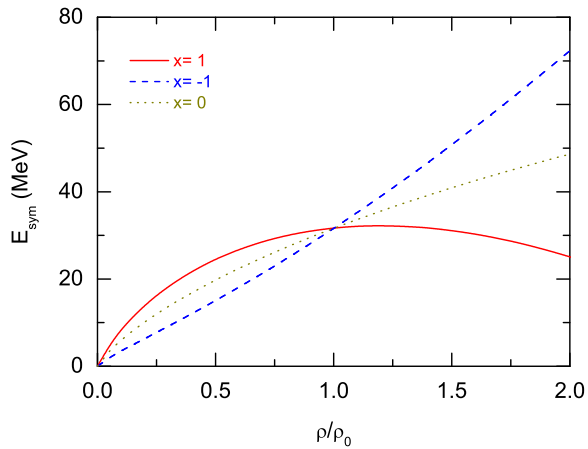}
\end{center}
\caption{(Color online) Density dependent nuclear symmetry
energy.} \label{esym}
\end{figure}
Another important input to the transport model is the mean field.
In the IBUU04 we use the momentum- and isospin-dependent single
nucleon potential (MDI) given in ref.\cite{das03}. In this
interaction a parameter $x$ was introduced to vary the density
dependence of the nuclear symmetry energy while keeping other
properties of the EOS fixed. Fig.\ \ref{esym} shows the density
dependent symmetry energy with $x=1$, $x=0$ and $x=-1$. Available
experimental data on isospin diffusion\cite{tsang04} and
isoscaling\cite{shetty07} have allowed us to constrain the
symmetry energy to be between the curves with $x=0$ and $x=-1$ at
subsaturation densities\cite{chen05,li05}. At high densities,
however, there is so far no experimental constraint available. One
of our major motivations here is to examine whether hard photons
can be used to constrain the symmetry energy at supranormal
densities as we were asked by the interested
experimentalists\cite{wie}.

\begin{figure}[th]
\begin{center}
\includegraphics[width=0.45\textwidth]{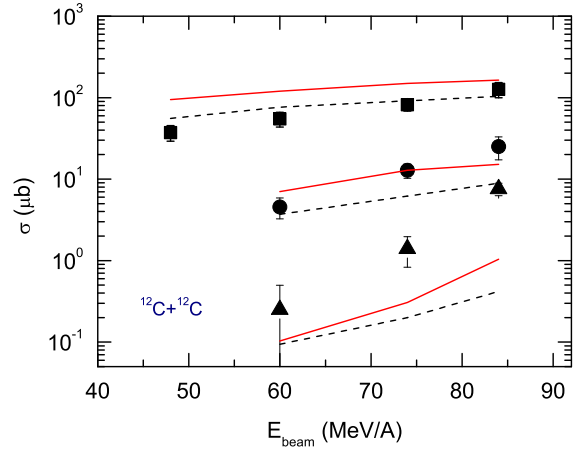}
\end{center}
\caption{(color online) Beam energy dependence of the inclusive
photon production cross sections in $^{12}$C+$^{12}$C collisions.
The solid symbols stand for experimental
data\cite{grosse86,cassrp}. (The squares are for 50 MeV $\leq
\varepsilon_{\gamma} <$ 100 MeV, circles for 100 MeV $\leq
\varepsilon_{\gamma} <$ 150 MeV and triangles for
$\varepsilon_{\gamma}\geq$ 150 MeV). The solid lines are
calculated using the $p^a_{\gamma}$ and the dashed ones using the
$p^b_{\gamma}$.} \label{com}
\end{figure}

While we are not aiming at reproducing any data in this
exploratory work, it is necessary to first gauge the model by
comparing with the available data. Shown in Fig.\ \ref{com} are
the calculations with both $p^a_{\gamma}$ and $p^b_{\gamma}$ and
the experimental data for the inclusive cross section of hard
photon production in the reaction of
$^{12}$C+$^{12}$C\cite{grosse86,cassrp}. The calculations are done
with $x=0$. It is seen that both calculations are in reasonable
agreement qualitatively with the experimental data except for the
very energetic photons. Quantitatively, the agreement is at about
the same level as previous calculations by others in the
literature\cite{ko85,bau86,gan94}. We notice that the uncertainty
in the elementary $pn\rightarrow pn\gamma$ probability leads to an
appreciable effect on the inclusive $\gamma$-production in
heavy-ion reactions. As one expects, the larger value of
$p^a_{\gamma}$ from the semi-classical picture gives significantly
higher $\gamma$-production cross section in heavy-ion reactions.
In fact, this effect is larger than the effects of the symmetry
energy obtained by varying the $x$ parameter from $x=0$ to $x=-1$
or $x=1$. It is thus a really very challenging task to extract
useful information about the symmetry energy from photon
production in heavy-ion reactions given the currently existing
uncertainties associated with the elementary probability.
Nevertheless, we are very hopeful that there are ways to overcome
this difficulty. As an example, we will actually discuss later in
this paper one possible way to do this.

\begin{figure}[th]
\begin{center}
\includegraphics[width=0.45\textwidth]{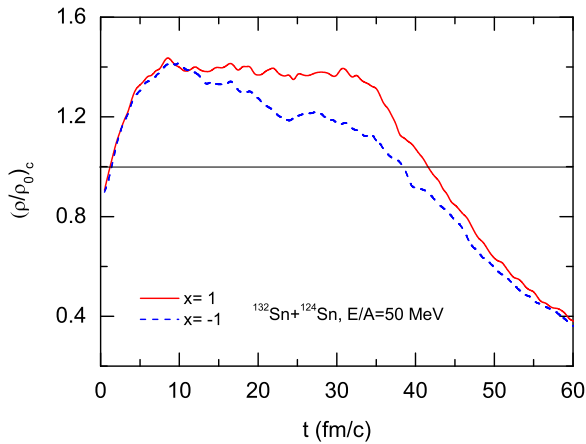}
\end{center}
\caption{(Color online) The central density reached in the head-on
reactions $^{132}$Sn+$^{124}$Sn at $50$ MeV/A with the symmetry
energies of x=1, -1 using the $p^b_{\gamma}$. The horizontal line
stands for the normal nuclear density.} \label{den}
\end{figure}

\begin{figure}[th]
\begin{center}
\includegraphics[width=0.45\textwidth]{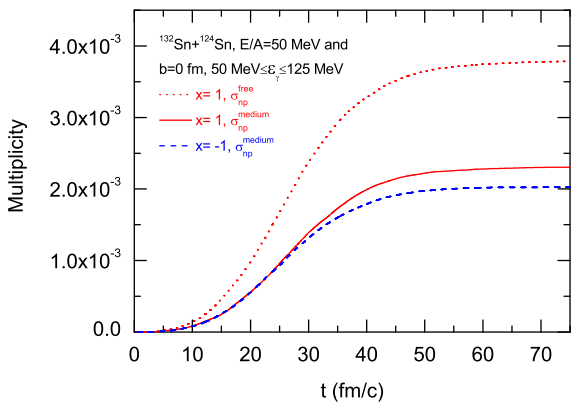}
\end{center}
\caption{(Color online) Time evolution of the multiplicity of hard
photons with 50 MeV $\leq \varepsilon_{\gamma} \leq$ 125 MeV in
the head-on collisions of $^{132}$Sn+$^{124}$Sn at a beam energy
of $50$ MeV/A using the $p^b_{\gamma}$.}\label{sp}
\end{figure}

We now turn to discussing effects of the symmetry energy on photon
production in more detail. As an example, we consider head-on
collisions of $^{132}$Sn+$^{124}$Sn at an incident energy of $50$
MeV/A using the symmetry energies of $x=1$ and $-1$. For this
discussion, it is sufficient to use only the $p^b_{\gamma}$.
First, we examine the evolution of the central density in Fig.\
\ref{den}. The highest central density reached is about
$1.3\rho_0$. At this density the symmetry energy with $x=-1$ is
higher by about 20\% as shown in Fig. \ref{esym}. The stiffer
symmetry energy with $x=-1$ leads to a slightly lower central
density. The supranormal density phase lasts from about 10 to 35
fm/c. It is in this phase, as shown in Fig.\ \ref{sp}, that most
of the hard photons with 50 MeV $\leq \varepsilon_{\gamma} \leq$
125 MeV are produced. This observation is consistent with
conclusions of the earlier
studies\cite{bertsch86,ko85,cassing86,bau86,stev86}. It is seen
that the softer symmetry energy with $x=1$ produce just a little
more hard photons. To investigate effects of the in-medium NN
cross sections, we also carried out calculations using the
free-space NN cross sections but with the same $p^b_{\gamma}$ for
the case of $x=1$. It is seen that the free-space cross sections
leads to significant higher hard photon production. This is simply
because the in-medium NN cross sections are smaller than the
free-space onces\cite{li05}. Moreover, we notice that the change
of the NN cross sections leads to a much larger effect than the
symmetry energy. But overall, both effects are very small compared
to that due to the uncertainty of the elementary
$\gamma$-production probability. Nevertheless, to put the
comparison into the proper context we stress that the symmetry
energy and the NN cross sections are only varied, respectively, by
at most 20\% and 50\% in the reaction considered by changing the
$x$ parameter between $1$ and $-1$ and the cross sections between
the free-space and in-medium ones. While the current uncertainty
of the $\gamma$-production probability is significantly larger.

\begin{figure}[th]
\begin{center}
\includegraphics[width=0.45\textwidth]{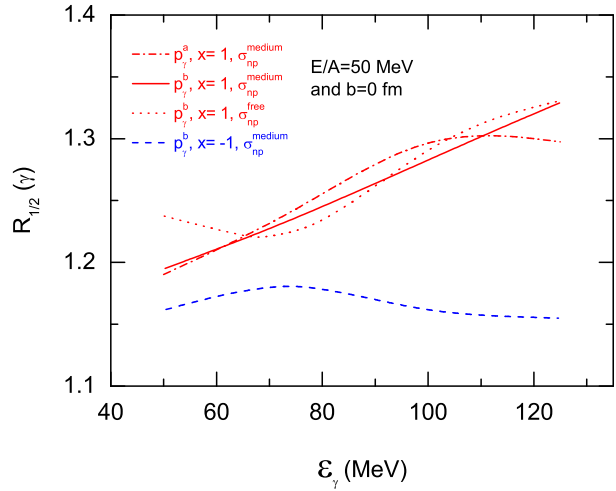}
\end{center}
\caption{(Color online) The spectra ratio of hard photons in the
reactions of $^{132}Sn+^{124}Sn$ and $^{112}Sn+^{112}Sn$ reactions
at a beam energy of $50$ MeV/A with the symmetry energies of x=1,
x=-1.} \label{relat}
\end{figure}

Given all of the uncertainties discussed above and the fact that
the cross section for hard photon production in intermediate
energy heavy-ion collisions is very small, is there anyway one can
reduce these uncertainties naturally and thus see more cleanly the
effects of the symmetry energy? Like in many experiments searching
for minute but interesting effects, ratios of two reactions can
often reduce not only the systematic errors but also some
``unwanted" effects. At least theoretically, within the
perturbative approach adopted here, the uncertainty of the
$\gamma$-production probability should get almost completely
cancelled out in the ratio of photons from two reactions. We thus
propose to measure experimentally the spectra ratio
$R_{1/2}(\gamma)$ of hard photons from the head-on reactions of
$^{132}$Sn+$^{124}$Sn and $^{112}$Sn+$^{112}$Sn, i.e.,
\begin{equation}
R_{1/2}(\gamma)\equiv\frac{\frac{dN}{d\varepsilon_{\gamma}}(^{132}Sn+^{124}Sn)}
{\frac{dN}{d\varepsilon_{\gamma}}(^{112}Sn+^{112}Sn)}.
\end{equation}
Depending on the relative number of neutron-proton scatterings in
the two reactions, uncertainties due to the NN cross sections can
also get significantly reduced. Shown in Fig.\ \ref{relat} is the
$R_{1/2}(\gamma)$ calculated for four cases using both the
$p^a_{\gamma}$ and $p^b_{\gamma}$. First of all, it is seen
clearly that the full calculations with the $p^a_{\gamma}$ and
$p^b_{\gamma}$ and the in-medium NN cross sections indeed lead to
about the same $R_{1/2}(\gamma)$ within statistical errors as
expected. It is also clearly seen that effect of the in-medium NN
cross sections get almost completely cancelled out. These
observations thus verify numerically the advantage of using the
$R_{1/2}(\gamma)$ as a robust probe of the symmetry energy
essentially free of the uncertainties associated with both the
elementary photon production and the NN cross sections. Most
interestingly, comparing the calculations with $x=1$ and $x=-1$
both using the $p^b_{\gamma}$, it is clearly seen that the
$R_{1/2}(\gamma)$ remains sensitive to the symmetry energy
especially for very energetic photons. Again, the symmetry energy
is varied by at most 20\% in the reaction considered by varying
the parameter $x$ from $1$ ro $-1$. Thus, the approximately 15\%
maximum change in the spectra ratio represents a relatively
significant sensitivity. To put things in perspective, we notice
that this sensitivity to the symmetry energy is actually at about
the same level as most hadronic probes including the $K^0/K^+$
ratio. The latter using strange particles is considered as among
the most clean hadronic probes of the symmetry
energy\cite{ditoro}. It shows about a 15\% change while the
symmetry energy is changed by at least 50\% at the density reached
in heavy-ion reactions near the kaon production
threshold\cite{ditoro}. Obviously, compared to the $K^0/K^+$ ratio
the hard photon production is an even more sensitive and clean
observable. Of course, we also notice that while photons are
completely free from final state strong interactions, besides
cosmic-radiation background, one needs to consider photons from
$\pi^{0}$ and fragment decays in the data analysis\cite{grosse86}.

Naturally, one may think about reactions at higher beam energies
to reach higher densities and thus to explore the behaviors of the
symmetry energy there. However, other channels for hard photon
productions may start playing more important roles at higher beam
energies. Moreover, the reaction dynamics will be then dominated
by nucleon-nucleon collisions rather than the nuclear mean-field.
Effects of the symmetry energy on photons are then expected to
become smaller. This is mainly because compared to nucleons that
are directly influenced by the symmetry potential, the hard
photons are affected by the symmetry potential only indirectly
through the momentum distributions and the densities of the
colliding proton-neutron pairs. This is also why the hard photons
were found not so sensitive to the nuclear equation of state in an
early study\cite{ko87}. Only at intermediate energies both the
mean-filed and the NN collisions play about equally important
roles in the reaction dynamics. In fact, we also calculated the
$R_{1/2}(\gamma)$ for the same reactions but at a beam energy of
$400$ MeV/A. We find a much small symmetry energy effect.

In summary, we have carried out an exploratory study about effects
of the symmetry energy on the production of hard photons from
intermediate energy heavy-ion reactions using a perturbative
approach within the IBUU04 transport model. Effects of the
symmetry energy on the yields and spectra of hard photons from
individual reactions are generally smaller than those due to the
existing uncertainties of both the in-medium nucleon-nucleon cross
sections and the elementary $pn\rightarrow pn\gamma$ probability.
Very interestingly, however, the ratio of hard photon spectra
$R_{1/2}(\gamma)$ from two reactions using isotopes of the same
element is not only approximately independent of these
uncertainties but also quite sensitive to the symmetry energy. The
sensitivity is at about the same level as most hadronic probes.
Compared to the $K^0/K^+$ ratio in heavy-ion reactions near the
kaon production threshold, hard photons are completely free of
final state strong interactions and are even more sensitive to the
symmetry energy.

B.A. Li would like to thank A. Chbihi, F. Gulminelli and J.P.
Wieleczko for their invitation and kind hospitality at Ganil and
LPC where this work got started after stimulating discussions with
them. We would like to thank Wei-Zhou Jiang, Che-Ming Ko and David
d'Enterria for helpful discussions. The work was supported in part
by the US National Science Foundation under Grant No. PHY-0652548,
the Research Corporation under Award No. 7123, the National
Natural Science Foundation of China under Grant Nos.
10575071,10675082, 10575119 and 10710172, MOE of China under
project NCET-05-0392, Shanghai Rising-Star Program under Grant No.
06QA14024, the SRF for ROCS, SEM of China, and the National Basic
Research Program of China (973 Program) under Contract No.
2007CB815004.

\end{document}